# Indices to Quantify the Ranking of Arabic Journals and Research Output

## Mahmoud Abdel-Aty

*Centre of Scientific Publications, University of Bahrain, 32038 Kingdom of Bahrain*
*Department of Mathematics, Sohag University, 82524 Sohag, Egypt*
*mabdelaty@uob.edu.bh*

**Abstract:** I propose two simple indices to classify journals, published in Arabic language, and different researchers. These indices depend upon the known impact factor and h-index. The new indices give an easy way to judge the rank of any journal (output of any researcher) without looking for other journals (output of other researchers).

Keywords: Impact Factor, H-index

## Introduction

The ranking of journals is an important issue for institutions and researchers around the world. The existing tools, specifically impact factor, allows users to configure their ranking interests, as well as provide a more reasonable method to evaluate a journal's impact. The impact factor, IF, is a measure of the frequency with which the average article in a journal has been cited in a particular year or period. It is one of the evaluation tools provided by Thomson Reuters Journal Citation Reports JCR [1-3].

The annual Journal Citation Reports impact factor is a ratio between citations and recent citable items published: a journal's impact factor is calculated by dividing the number of citations of the current year to the source items published in that journal during the previous two years. The IF is used to compare different journals within a certain field. In a given year, the impact factor of a journal is the average number of citations received per paper published in that journal during the two preceding years [2]. For example, if a journal has an impact factor of 5 in 2012, then its papers published in 2010 and 2011 received 5 citations each on average in 2012. The 2012 impact factor of a journal would be calculated as follows:

$$IF = A/B,$$

where $A$ the number of times articles published in 2010 and 2011 were cited by indexed journals during 2012 and B is the total number of "citable items" published by that journal in 2010 and 2011. ("Citable items" are usually articles, reviews, proceedings or notes; not editorials or Letters-to-the-Editor.)

New journals, which are indexed from their first published issue, receive an impact factor after two years of indexing; in this case the citations to the year prior to Volume 1 and the number of articles published in the year prior to Volume 1 are known zero values. Journals that are indexed starting with a volume other than the first volume do not receive an impact factor until they have been indexed for three years. Annuals and other irregular publications sometimes publish no items in a particular year and this affects the count. The impact factor relates to a specific time period; it is possible to calculate it for any desired period and the Journal Citation Reports (JCR) also includes a 5-year impact factor [3].

In 2005, J. E. Hirsch [4] proposed an index to quantify an individual's scientific research output which is called the h-index. Hirsch defined his index as the number of papers with citation number higher or equal to h, i.e., a scientist has index $h$ if $h$ of his/her $N_p$ papers have at least $h$ citations each, and the other ($N_p$-$h$) papers have no more than $h$ citations each. This index is used to characterize the scientific output of a researcher in a very good way. The way of calculation h-index includes the total number of papers published over a certain period of years and the number of citations for each paper. Now the ISI (Institute of Scientific Information) Web of Knowledge indexes more than 11,000 science and social science journals [3] and uses the impact factor to report on ranking the journals. Also Scopus [5] uses the h-index to characterize the scientific output of a researcher.

Leo Egghe [6] has suggested what is called *g*-index. This index is used to quantify scientific productivity based on publication record and is calculated based on the distribution of citations received by a given researcher's publications. Given a set of articles ranked in decreasing order of the number of citations that they received, the *g*-index is the (unique) largest number such that the top *g* articles received (together) at least $g^2$ citations. In agreement with *h*-index, the *g*-index is a number close to *h*-index for the same authors.

In this communication I introduce two alternate indices that can be used to estimate of the impact of Journals published in Arabic Language as well as scientists' cumulative research contributions.

**Suggested Model**

The above methods have been motivating the investigation of possible models to be applied to journals published in Arabic and also researchers who write in Arabic thereby avoiding the complications related to the use of the impact factor and h-index. On one hand there is the question of what kind of indications the impact factor (*h*-index) means. This question has already been answered [5] and it turns out to be a good indicator if we know the impact factor (*h*-index) of other journals, i.e., this number has no meaning if we do not know the other journals (researchers) impact factor (*h*-index). From this fact it follows that, in order to achieve the strongest indicators for any journal or researcher, one should think of building a model system capable of creating incoming percentage value.

For a journal's rank I suggest to use the following model

$$AF = \left(\frac{A}{A+B}\right) \times 100,$$

where *A* is the total number of times articles published in this journal in a preceding year were cited by indexed journals during the following year and B is the total number of "citable items" published by that journal in the preceding year. The application of this model shows the results corresponding to the impact factor, but here we can clearly know the impact of any journal without looking for the impact of other journals, since the result is a percentage between *0* and *100*.

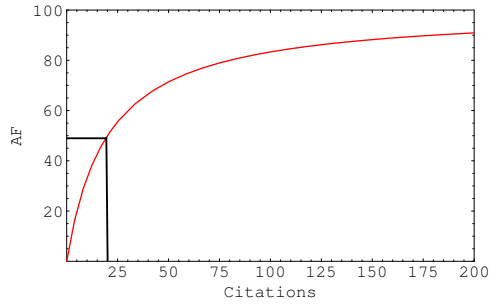

Fig. 1: Plot of *AF* as a function of the number of citations. The number of papers published in the preceding year was assumed to be 20.

In Fig. 1 we plot the relation between *AF* and the number of citations. We fix the number of published papers as *20* papers in the preceding year. It is shown that, as the number of citations is increased, the *AF* increases in agreement with the *IF*. For example, if the number of citations is *5,* then the AF is *20%*. It is interesting to note that, when the number of citations is increased dramatically compared with the number of published papers in the preceding year, the *AF* increases by very small values. This means that this new factor is very sensitive when the number of published papers in preceding year of a journal is close to or less than the number of citations.

*Table 1*

| Citations | 5 | 10 | 20 | 40 |
|---|---|---|---|---|
| AF | 20 | 33.3 | 50 | 66.6 |

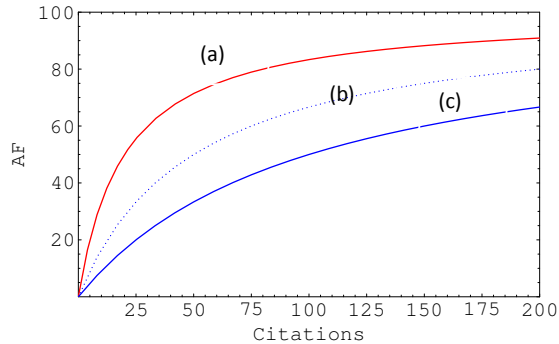

Fig. 2: Example of three different journals with different numbers of published papers (N) in the presiding year, where (a) N=20, (b) N=50 and (c) N=100.

From Fig. 2 it is shown that *AF* decreases if the number of published papers in a journal is increased with a fixed value of the number of citations.

For a given individual I suggest to use a modified version of the Hirsch index as follows

$$AsF = \left(\frac{h}{h+1}\right) \times 100.$$

From this equation it is shown that *AsF* is an increasing number according to the increasing of h, but does not exceed 100. In particular, assume that the researcher has *h*-index *1*. This corresponds to *AsF=50%* and h-index *5* corresponds to *AsF=83.3\%* and so on. When the *h*-index gets larger, the difference in *AsF* became smaller (see table 2)

*Table 2*

| *h*-index | *1* | *5* | *10* | *20* | *40* | *100* |
|---|---|---|---|---|---|---|
| *AsF* | *50* | *83.3* | *90.9* | *96.2* | *97.6* | *99* |

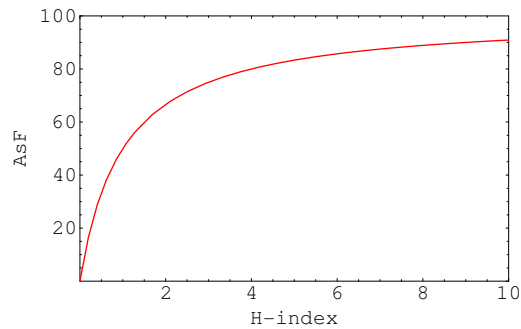

**Fig. 3:** Plot of *AsF* as a function of the *h*-index.

**In conclusion**, we have shown that the suggested AF independently gives a direct indication of any journal's rank. Also the suggested model, *AsF*, for estimating the impact of the scientist's cumulative research contributions gives a direct percentage that can be used to compare different individuals. Finally we showed that the simultaneous growth of *AF* (*AsF*) is dependent upon the growth of *IF* (*h*-index).

**Acknowledgments:** I thank, Peter Leach, Arthur Mc-Gurn, A.-S. F. Obada, A Al-Hasheme, K. Al-Khalili and W. Alnaser for helpful discussions. Also I would like to thank the referees for constructive comments.